\documentclass{iopbk2e}
\usepackage{epsf}

\def\ltsima{$\; \buildrel < \over \sim \;$}
\def\lsim{\lower.5ex\hbox{\ltsima}}
\def\gtsima{$\; \buildrel > \over \sim \;$}
\def\gsim{\lower.5ex\hbox{\gtsima}}
\def\hmpc{\ifmmode\,h^{-1}\,{\rm Mpc}\;\else$h^{-1}\, {\rm Mpc}\;$\fi}
\def\kmpc{\ifmmode\,h\,{\rm Mpc^{-1}}\;\else$h\,$Mpc^{-1}\;\fi}
\def\x{{$\xi(r)$\ }}

\def\kms{\,{\rm km\,s^{-1}}}

\def\uua{{\bf u}_1\ }
\def\uub{{\bf u}_2\ }

\def\bfv{{\bf d}}

\def\ro{r_\circ}
\def\n_med{{\left<n\right>}}

\def\begc{\begin{center} }
\def\endc{\end{center} } 
\def\begf{\begin{figure} }
\def\endf{\end{figure} }

\def\apj{{\it Astrophys. J.\/}~}
\def\mnras{{\it Mon. Not. R. Astr. Soc.}~}
\def\aa{{\it Astron. Astrophys.}~}
\def\aj{{\it Astron. J.\/}~}
\def\nat{{\it Nature}~}
\begin{document}

\pagenumbering{arabic}
\setcounter{page}{1}

\Chapter[From Clustering to Homogeneity]{Clustering in the Universe:
from Highly Nonlinear Structures to Homogeneity}{L
Guzzo\\Osservatorio Astronomico di Brera}

\section{Introduction}
This chapter\footnote{Lectures delivered
at the {\it Graduate School in Contemporary Relativity and
Gravitational Physics}, Como, May 2000} concentrates on a few specific topics
concerning the distribution of galaxies on scales from 0.1 to nearly
1000 \hmpc. The main aim is to provide the reader with the 
information and tools to familiarize with a few basic questions: 1) What are
the scaling laws 
followed by the clustering of luminous objects over almost four decades
of scales; 2) How galaxy motions distort the observed maps in redshift
space, and how we can correct and use them to our benefit; 3) Is the observed
clustering of galaxies suggestive of a fractal Universe; and
consequently, 4) Is our faith in the Cosmological 
Principle still well placed, i.e. do we see evidence for a homogeneous
distribution of matter on the largest explorable scales, in terms of
the correlation function and power spectrum of the distribution of
luminous objects. For some of these questions we have a well--defined
answer, but for some others the idea is to indicate the path along
which there is still a good deal of exciting work to be done.
  
\section{The clustering of galaxies}
\label{sec:xi}

\begin{figure}
\centering
\epsfysize=16cm 
  \hspace{1.3cm}
\epsfbox{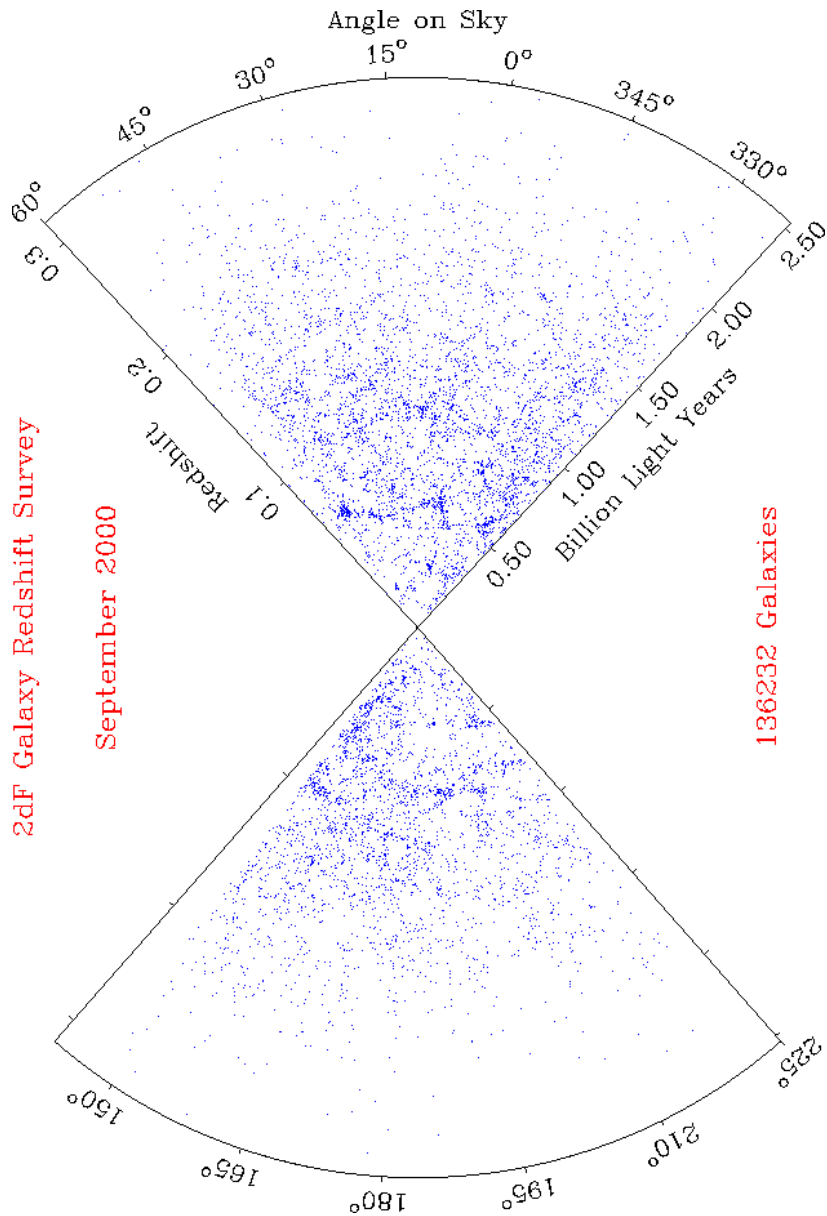} 
\caption{The distribution of the nearly 140,000 galaxies observed so far
(September 2000) in the 2dF survey (from \cite{2dF_www}): 
compare this picture to that in \cite{Texas} to see how rapidly this
survey is progressing towards its goal of 250,000 redshifts measured
(note that this is a projection over a variable depth in declination,
due to the survey being still incomplete).  
}
\label{2dF-cone}
\end{figure}

I believe most of the students attending this School are familiar with 
the beautiful {\sl cone diagrams} \index{cone diagrams} showing the
distribution of galaxies in what have been often called {\sl slices of
the Universe}.  This has 
been made possible by the tremendous progress in the efficiency of
\index{redshift surveys} redshift surveys, 
i.e. observational campaigns aimed at measuring the distance of large
samples of galaxies through the cosmological redshift observed in
their spectra.   This is one of the very simple, yet fundamental
pillars of observational cosmology: reconstructing the
three--dimensional positions of galaxies in space to be able to
study and characterise statistically their distribution. 
Figure~\ref{2dF-cone} shows the current status of the
ongoing 2dF survey \index{2dF survey} and gives an idea of the state
of the art, 
with $\sim 130,000$ redshifts measured and a planned final number of
250,000 \cite{2dF_Dunk}.  From this plot, the main features of the
galaxy distribution can be appreciated.  One can easily recognize {\sl
clusters}, {\sl superclusters } \index{superclusters} and
\index{voids} {\sl voids}, and
get the feeling on how the galaxy   
distribution is extremely inhomogeneous to at least 50 \hmpc (see 
\cite{Texas} for a more comprehensive review).
%


The inhomogeneity we clearly see in the galaxy distribution can be
quantified at the simplest level by asking what is the {\sl excess}
probability over random to find a galaxy at a separation 
$r$ from another one.  This is one way by which one can define the
{\sl two--point correlation function}, \index{correlation function}
certainly the most 
perused statistical estimator in Cosmology (see \cite{Peebles80} for a more
detailed introduction).   When we have a catalogue with only galaxy
positions on the sky (and usually their magnitudes), however, the first 
quantity we can compute is the {\sl angular} correlation function
\index{angular correlation function}
$w(\theta)$.  This is a projection of the {\sl spatial} correlation function
$\xi(r)$ along the redshift path covered by the sample.  The relation
between the angular and spatial functions is expressed for small
angles by the
{\sl Limber equation} (see \cite{JAP} and \cite{Peebles80} for
definitions and details)
\begin{equation}
w(\theta) = \int_0^\infty dv\, v^4\, \phi^2(v) \int_{-\infty}^\infty du\,
\xi\left(\sqrt{u^2+v^2\theta^2}\right) \,\,\,\,\,   ,
\end{equation}
where $\phi(v)$ is the {\sl radial selection function} of the 
two--dimensional catalogue, that in this version gives the comoving
density of objects at a given distance $v$ (which depends, for 
example, on the magnitude limit of the catalogue and the specific
luminosity function of the type of galaxies one is studying).
For optically--selected galaxies \cite{APM90,EDSGC} $w(\theta)$ is well
described by a power--law shape $\propto 
\theta^{-0.8}$, corresponding to a spatial correlation function
$(r/r_o)^{\gamma}$, with $r_o\simeq 5\hmpc$ and $\gamma\simeq -1.8$,
and a break with a rapid
decline to zero around scales corresponding to $r\sim 30\hmpc$.  

\begin{figure}
\epsfxsize=11cm 
  \hspace{0.2cm}
\epsfbox{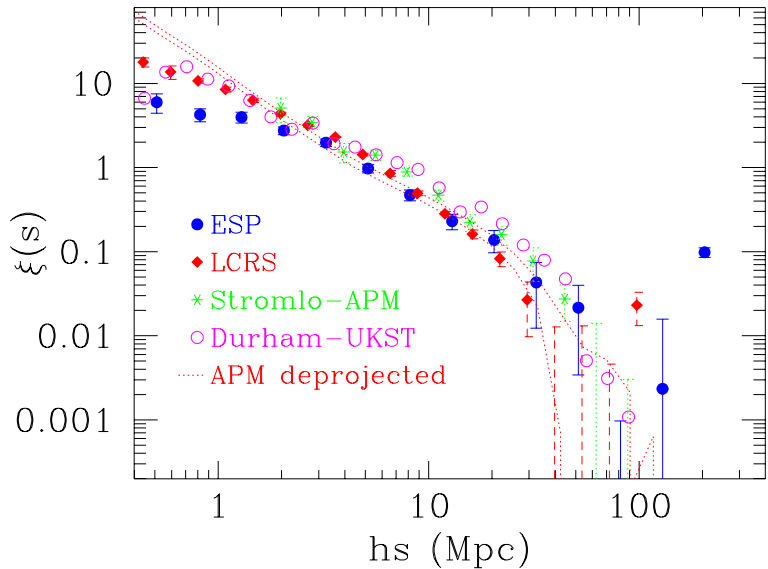} 
\caption{The two--point correlation function of galaxies, as measured
from a few representative optically--selected surveys (from
\cite{Texas}).  The plot shows results from the ESP \cite{GuzzoESP},
LCRS \cite{Tucker}, APM-Stromlo, \cite{Loveday92b} and Durham-UKST
\cite{DUKST_xi} surveys, plus the 
real--space $\xi(r)$ de--projected from the angular 
correlation function $w(\theta)$ of the APM survey \cite{Baugh96}.}
\label{xi-surveys}
\end{figure}

The advantage of angular catalogues still at present is the large
number of galaxies they include, up to a few millions \cite{APM90}.
Since the beginning of the eighties (e.g. \cite{DP83}),  
redshift surveys allowed us to compute directly $\xi(r)$ in
three--dimensional space, and the most recent samples have pushed these
estimates to separations of $\sim 100$ \hmpc (e.g. \cite{GuzzoESP}).
Figure~\ref{xi-surveys} shows the two--point correlation function in
{\sl redshift space}\footnote{This means that distances are
computed from the red--shift in the galaxy spectrum,
neglecting the Doppler contribution by its peculiar velocity which
adds to the Hubble flow (\S~\ref{sec:distort})}, indicated as $\xi(s)$, for a
representative set of  
published redshift surveys \cite{GuzzoESP, Tucker, Loveday92b,
DUKST_xi}.   In addition, the dotted lines show the real--space 
$\xi(r)$ obtained through de--projection of the angular $w(\theta)$ from
the APM galaxy catalogue \cite{Baugh96}.  The two different lines
correspond to two different assumptions about galaxy clustering
evolution, which has to be taken into account in the de--projection,
given the depth of the APM survey.  This illustrates some of the
uncertainties inherent in the use of the angular function. 
As can be seen from the Figure, the shape of $\xi(s)$ below $5-10\hmpc$ is
reasonably well described by a power law, but for the four redshift
samples the
slope is shallower than the canonical $\sim -1.8$ nicely followed by the APM
$\xi(r)$.  This is 
due to the redshift--space smearing of structures that suppresses the
true clustering power on small scales, as we shall discuss in the
following section.   Note how $\xi(s)$ maintains a low--amplitude,
positive value out to separations 
of more than $50\hmpc$, showing explicitly why large--size galaxy 
surveys are important: we need large volumes and good statistics to be
able to extract such a weak clustering signal from the noise.
Finally, the careful reader might have noticed a small but significant 
positive change in the slope of the APM $\xi(r)$ (the only one for which
we can see the 
undistorted real--space clustering at small separations), around $r\sim
3-4\hmpc$.  On scales larger than this, all data show a ``shoulder''
before breaking down.  This inflection point appears around 
the scales where $\xi\sim 1$, thus suggesting a relationship with the
transition from the linear regime (where each mode of the power
spectrum grows by the same amount and the shape is preserved), 
 to fully nonlinear clustering on smaller scales \cite{G91}.  We 
shall come back to this in \S\ref{sec:scaling}.

\section{Our distorted view of the galaxy distribution}
\label{sec:distort}
\begin{figure}
\centering
\epsfysize=12cm 
\vspace{-3cm}
  \hspace{0cm}
\epsfbox{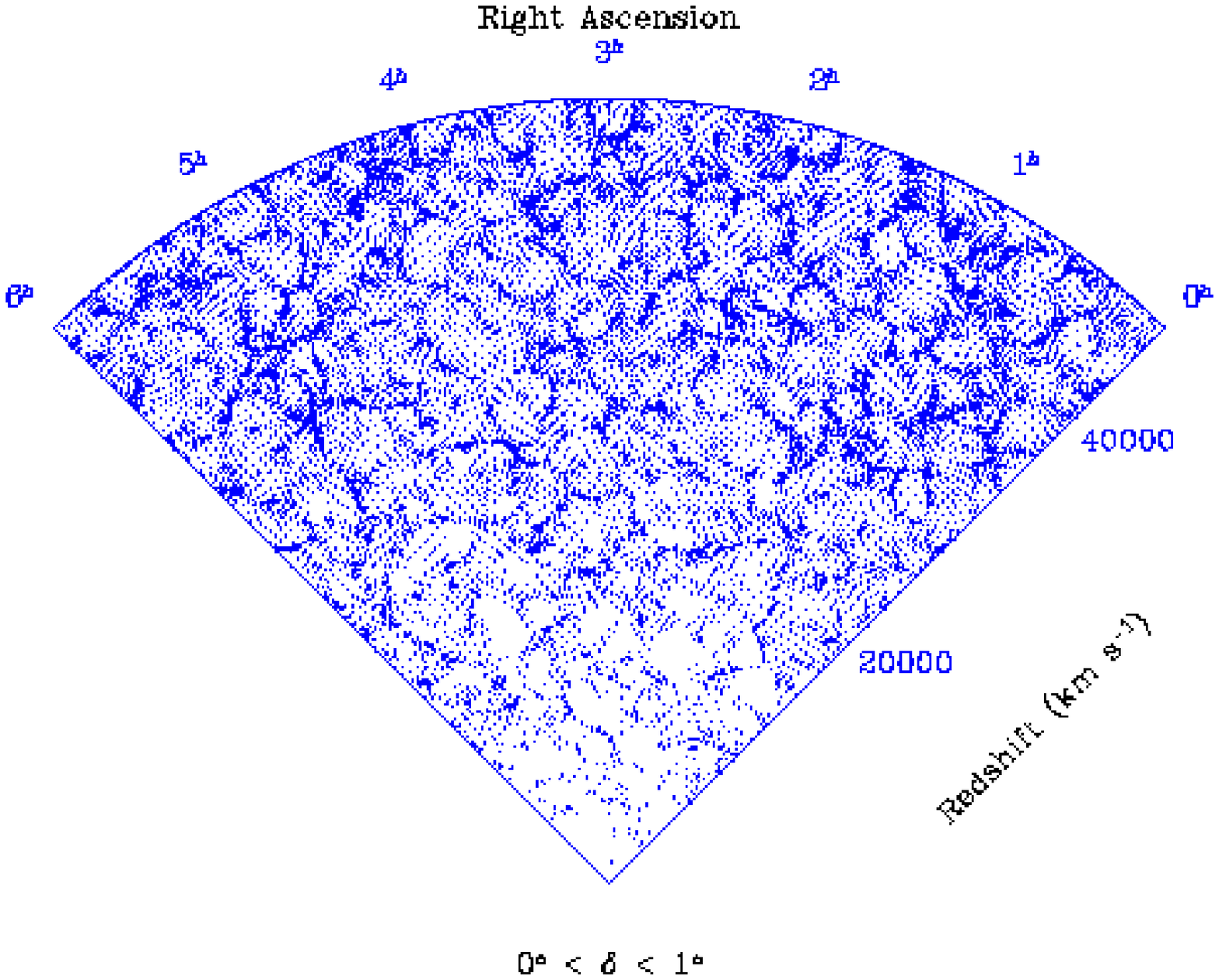} 
%
%
\centering
\epsfysize=12cm 
  \vspace{-2.1cm}
  \hspace{0cm}
\epsfbox{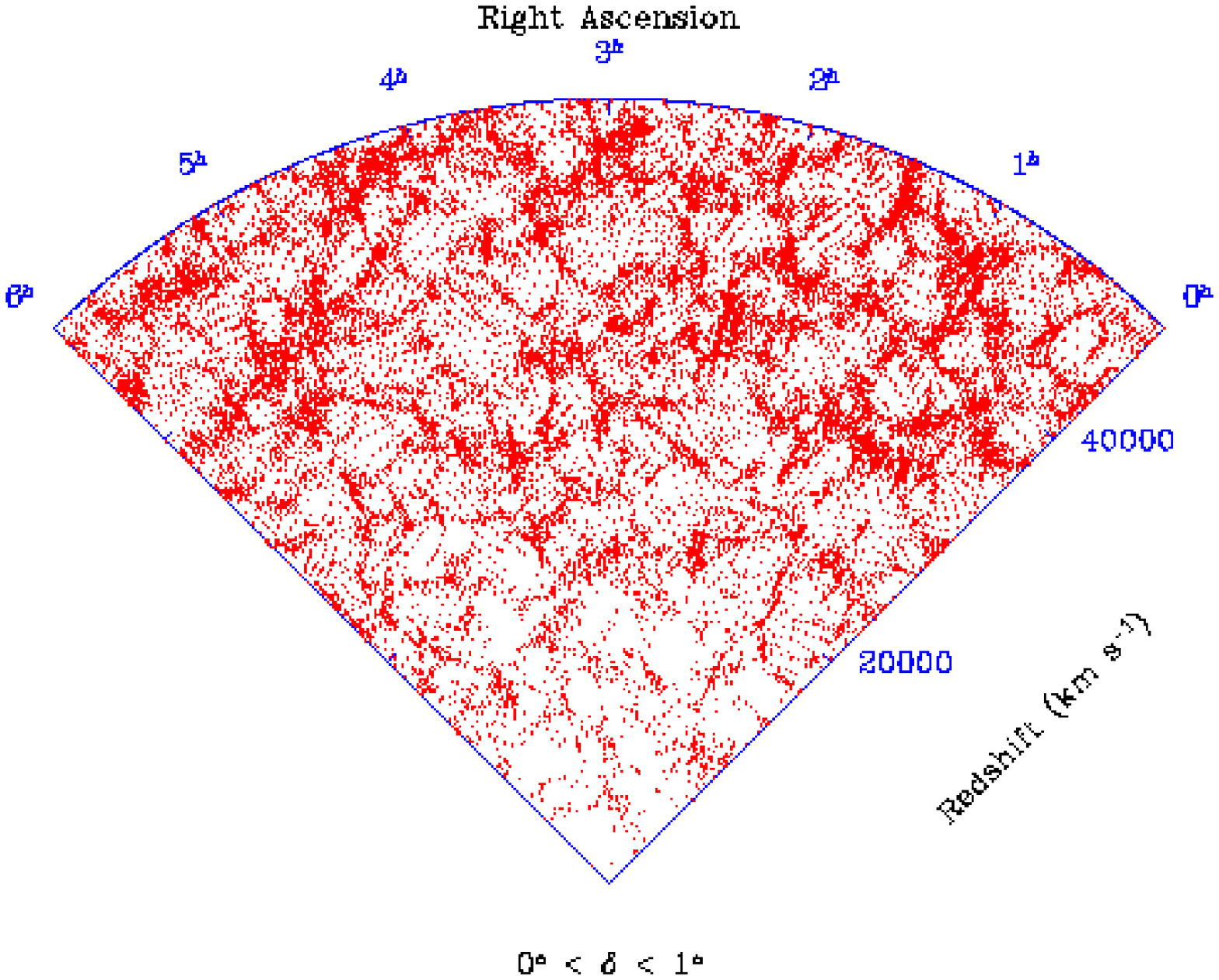} 
\caption{Particle distribution from a 1--degree thick mock survey
through a large--size Tilted--CDM n-body 
simulation in real (top) and redshift space (bottom).  The appearance
of the two diagrams gives a clear visual impression of the effect of
redshift--space distortions (note that here, unlike in the real survey 
of Figure~\ref{2dF-cone}, no apparent luminosity selection is applied, 
i.e. the sample is {\sl volume limited}).}
\label{sim-cones}
\end{figure}

\index{Redshift space distortions}
We just had an explicit example of how unveiling the true scaling laws
describing galaxy clustering from redshift surveys is complicated by
the effects of galaxy peculiar velocities. 
Separations between galaxies 
-- indicated as 
$s$ to remark this very point -- are not measured in real
3D space, but in 
{\sl redshift space}: what we actually measure when we take the
redshift of a galaxy is the quantity $cz=cz_{\rm true}+v_{\rm pec//}$,
where $v_{\rm pec//}$ is the component of the galaxy peculiar velocity
along the line of sight.  This quantity, while being typically $\sim 100
\kms$ for ``field'' galaxies, can rise above $1000 \kms$ in rich
clusters of galaxies.  As explicitly visible in
Figure~\ref{xi-surveys}, the resulting $\xi(s)$ is {\sl flatter}
than its real--space counterpart.  This is the result of two concurrent
effects: on small scales, clustering is suppressed by high
velocities in clusters of galaxies, that spread close pairs along the
line of sight producing in redshift maps what are sometimes called
``Fingers of God''.  Many of these are recognisable in
Figure~\ref{2dF-cone} as thin radial structures, particularly in the
denser part of the upper cone.  The net effect on $\xi(s)$ is in fact to
suppress its amplitude below $\sim 1-2 \hmpc$.  On the other hand, on
larger scales where motions are still 
coherent, streaming flows towards higher--density structures
enhance their apparent contrast when they appear to lie
perpendicularly to the line of sight. This, on the contrary, amplifies
$\xi(s)$ above $10-20\hmpc$.  
Both effects can be better appreciated with the help of a
computer N--body simulation, for which we have the leisure to see both
a real-- and a redshift--space snapshot, as in Figure~\ref{sim-cones}. 
\begin{figure}
\centering
\epsfxsize=8cm 
  \hspace{0cm}
\epsfbox{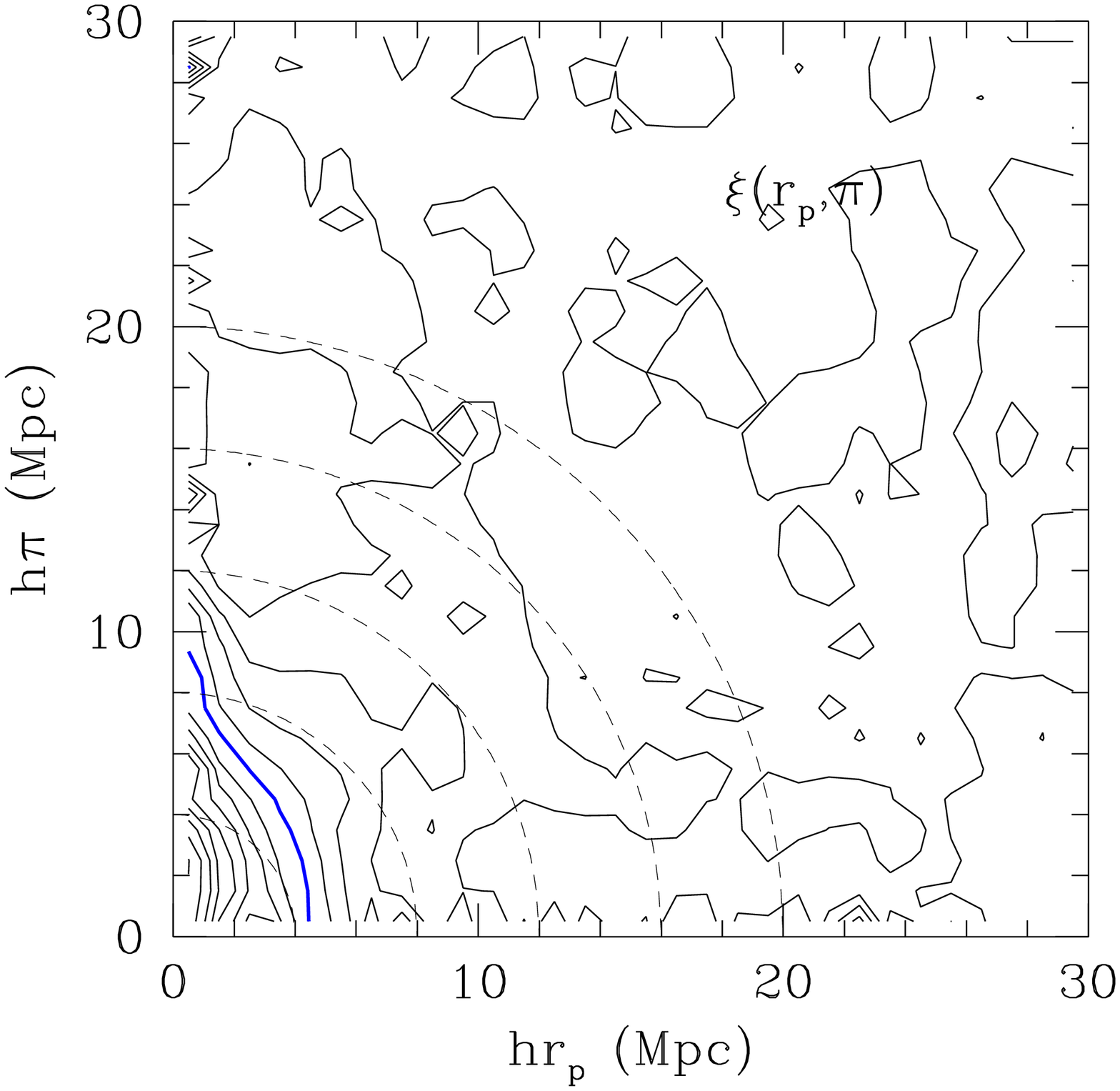} 
\caption{The typical appearance of the bidimensional correlation
function $\xi(r_p,\pi)$, in this specific case computed for the ESP survey
\cite{GuzzoESP}.  Note the elongation of the contours along the $\pi$
direction for small values of $r_p$, produced by high--velocity pairs
in clusters.  The dashed circles show contours of equal correlation in 
the absence of distortions.}  
\label{xi_rp_pi}
\end{figure}

How could we recover the correlation function of the undistorted
spatial pattern, i.e. $\xi(r)$?   This can be accomplished by computing
the two--dimensional correlation function
$\xi(r_p,\pi)$, where the radial separation $s$ of a galaxy pair
is split into two components, $\pi$, parallel to the line of sight,
and $r_p$, perpendicular to it, defined as follows
\cite{Fisher94_xi}. If $\bfv_1$ and $\bfv_2$ are the distances to the
two objects (properly computed) and we define the line of
sight vector ${\bf l} \equiv 
(\bfv_1 + \bfv_2)/2$ and the redshift difference vector ${\bf s}
\equiv \bfv_1 - \bfv_2$, then one defines
\begin{equation}
\pi \equiv {{{\bf s} \cdot {\bf l}}\over{|{\bf l}|}}\quad\quad
r_p^2 \equiv {{\bf s} \cdot {\bf s}} - \pi^2\,\,\, .
\label{f94-def}
\end{equation}
The resulting correlation function is a
bidimensional map, whose contours at constant correlation look as
in the example of Figure~\ref{xi_rp_pi}. By projecting $\xi(r_p,\pi)$ along the $\pi$
direction, we obtain a function that is independent of the distortion,
\begin{equation}
w_p(r_p) \equiv 2 \int_0^{\infty} d\pi\, \xi(r_p,\pi) = 
2 \int_0^{\infty} dy \, \xi_R\left[(r_p^2 + y^2)^{1/2}\right]\, ,
\label{wp}
\end{equation}
and is directly related to the real--space correlation function (here
indicated with $\xi_R(r)$ for clarity), as shown.  Modelling 
$\xi_R(r)$ as a power law, $\xi_R(r) = (r/r_0)^{-\gamma}$ we can
carry out the integral analytically, yielding
\begin{equation}
w_p(r_p)=r_p\left({r_0\over r_p}\right)^\gamma {\Gamma({1\over 2})\,
\Gamma({\gamma-1\over 2}) \over \Gamma({\gamma\over 2})}
\end{equation}
where $\Gamma$ is the Gamma function.   Such form can then
be fitted to the observed $w_p(r_p)$ to recover the parameters describing
$\xi(r)$ (e.g. \cite{G97}).  Alternatively, one can perform a formal Abel
inversion of $w_p(r_p)$ \cite{Ratcliffe}.

So far, we have treated redshift--space distortions nearly as only an 
annoying feature that prevents the true distribution of galaxies to be
directly seen.  In fact, being a dynamical effect they carry precious
direct information on the distribution of mass, independently
from the distribution of luminous matter.  This information can be
extracted, in particular by measuring the value of the {\sl pairwise
velocity dispersion } $\sigma_{12}(r)$. \index{pairwise velocity dispersion}
This is in practice a measure of the small--scale ``temperature'' of the
galaxy soup, i.e. the amount of kinetic energy produced by
the differences in the potential energy created by density
fluctuations.  Thus, finally, a measure of the mass variance on small scales.

$\xi(r_p,\pi)$ can be modelled as the convolution of the
real--space correlation function with the distribution function of
pairwise velocities along the line of sight \cite{DP83, Fisher_vel},
Let $F({\bf w},r)$ be the distribution function of 
the vectorial velocity differences ${\bf w} = \uub - \uua$ for pairs 
of galaxies separated by a distance $r$ (so a function of four
variables, $w_1$, $w_2$, $w_3$, $r$).   Let $w_3$ be the component of {\bf w}
along the direction of the line of sight (that defined by ${\bf l}$);
we can then 
consider the corresponding distribution function of $w_3$, 
\begin{equation}
f(w_3,r) = \int dw_1\, dw_2\, F({\bf w}, r) \,\,\,\, ,
\end{equation}
It is this distribution function that is convolved with $\xi(r)$
to produce the observed $\xi(r_p,\pi)$.  If we now call $y$ the component of the 
separation $r$ along the line of sight, with our convention we have that 
$w_3 = H_o(\pi - y)$ and the convolution
\begin{equation}
1+\xi(r_p,\pi) = \left[1+\xi(r)\right] \otimes f(w_3,r) \,\,\,\, ,
\end{equation}
can be expressed as
\begin{equation}
1+\xi(r_p,\pi) = H_\circ \int_{-\infty}^{+\infty} dy \left\{1+\xi\left[
(r_p^2 + y^2)^{1\over 2} \right]\right\}\,f\left[H_\circ (\pi - y)\right] 
\,\,\, .
\label{xp_model}
\end{equation} 
Note that this expression gives essentially a model description of the {\it
effect} produced by peculiar motions on the observed correlations, but does
not take into account the intimate relation between the mass density
distribution and the velocity field which is in fact a product of mass 
correlations (see \cite{Fisher95} and \cite{Sheth} and references
therein).  Within this model, therefore, 
we have no specific physical reason for choosing one or another form for the 
distribution function $f$.  Peebles \cite{Peebles_expmodel}
first showed that an exponential distribution best fits the observed data,
a result subsequently confirmed by N--body models \cite{Zurek94}.
According to this choice, $f$ can then be parameterised as
\begin{equation}
f(w_3,r) = {1\over\sqrt{2} \sigma_{12}(r)} \exp\left[{-\sqrt{2}\left|{w_3(r) - 
\langle w_3(r) \rangle \over \sigma_{12}(r)} \right|}\right]\,\,\, ,
\end{equation}
where $\langle w_3(r) \rangle$ and $\sigma_{12}(r)$ are respectively the first 
and second moment of $f$.  The projected {\sl mean streaming}
\index{mean streaming}$\langle w_3(r) 
\rangle$ is usually explicitly expressed in terms of $v_{12}(r)$, the 
first moment of the distribution $F$ defined above, i.e. {\sl the mean 
relative velocity of galaxy pairs with separation $r$}, 
$\langle w_3(r)\rangle = y\,v_{12}(r)/r$.  The final expression for
$f$ becomes therefore
\begin{equation}
 f(w_3,r) = {1\over \sqrt{2} \sigma_{12}(r)}\,\exp{\left\{ -\sqrt{2}H_0
 \left|{\pi - y\left[1+{v_{12}(r) \over H_0 r}\right] \over
\sigma_{12}(r)}\right| \right\}} \,\,\, , 
\label{f_v}
\end{equation} 
(see e.g. \cite{Fisher_vel} and \cite{G97} for more details). 

\begin{figure}
\centering
\epsfxsize=5cm 
  \hspace{0cm}
\epsfbox{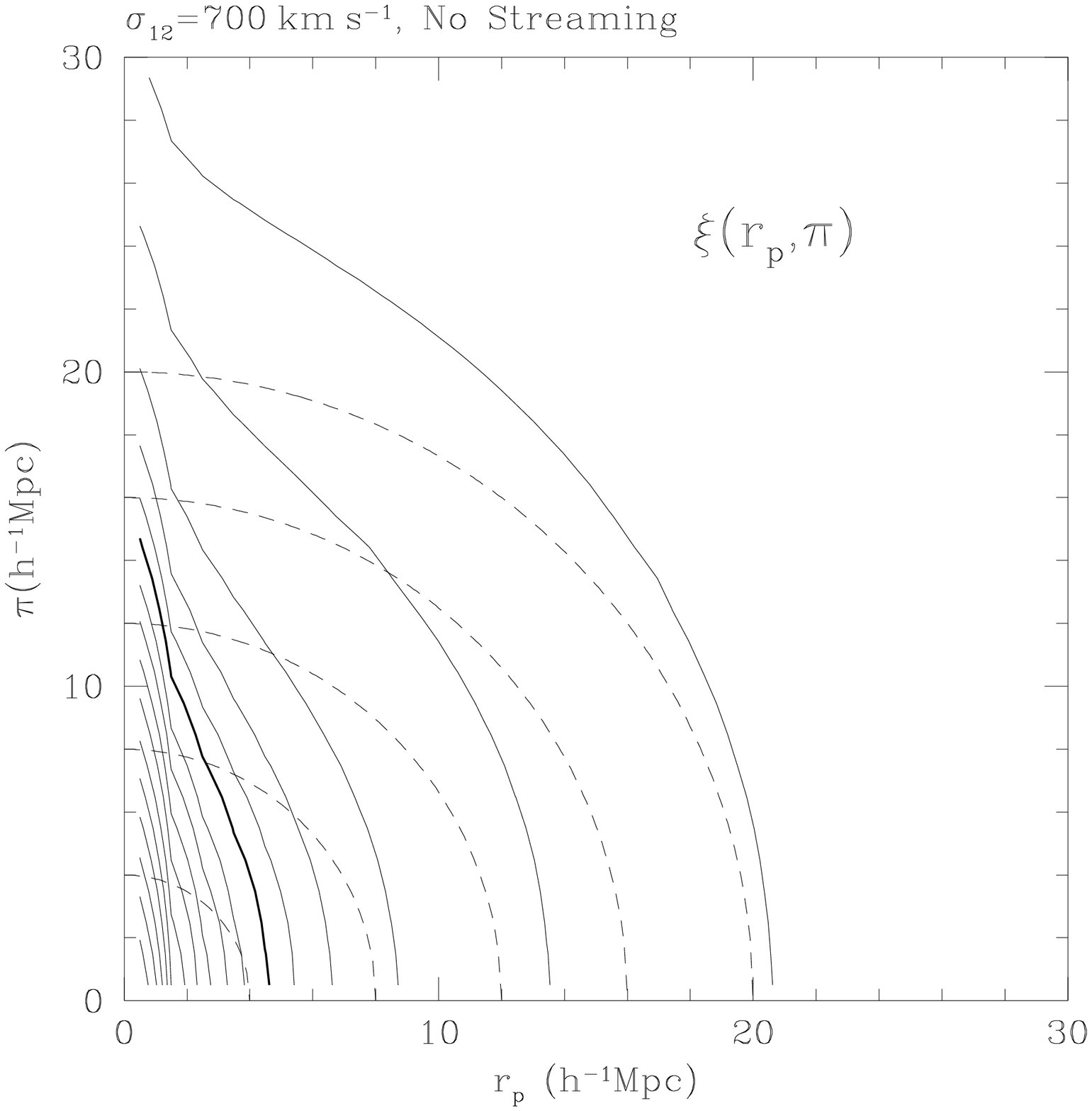} 
%
%
\centering
\epsfxsize=5cm 
  \hspace{0cm}
\epsfbox{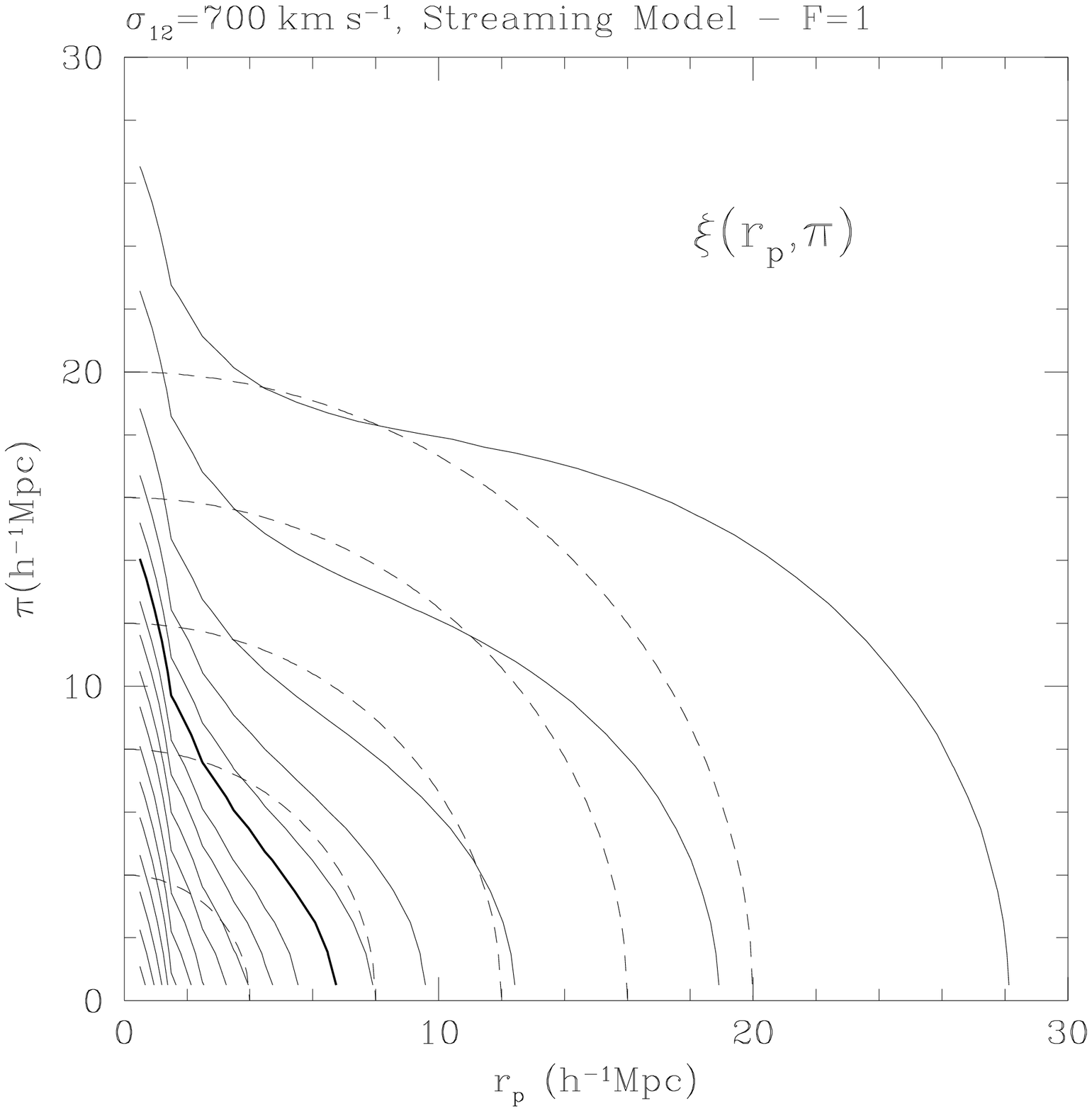} 
\caption{The relative effect of the mean streaming $v_{12}(r)$ and
pairwise velocity dispersion $\sigma_{12}(r)$ on the shape of the
contours of $\xi(r_p,\pi)$, seen through the model of eq.~\ref{xp_model}.  While
a high pairwise dispersion, $\sigma_{12}=700\kms$ independent on scale
is assumed (a reasonable approximation), the two cases of zero mean
streaming ($F=0$) 
and stable clustering ($F=1$) are considered in the infall model of
Davis \& Peebles \cite{DP83}.  Here the effect of the coherent motions
is more evident than in the data plot of Figure~\ref{xi_rp_pi}: the
contours of $\xi(r_p,\pi)$ are clearly compressed along the $\pi$ direction.
This compression is a measure of $\Omega_M^{0.6}/b$.}  
\label{xp-models}
\end{figure}
The practical estimate of $\sigma_{12}(r)$ is typically performed on the
data by fitting the model of eq.~\ref{xp_model} to a cut at fixed $r_p$ of the
observed $\xi(r_p,\pi)$.  To do this, one has first to estimate $\xi(r)$ from the
projected function $w_p(r_p)$ and choose a model for the 
mean streaming $v_{12}(r)$, as e.g. that based on the similarity solution
of the BBGKY equations \cite{DP83}:  
\begin{equation}
v_{12}(r) = -H_0 r {F \over 1+\left({r\over r_0}\right)^2} \,\,\, .
\label{eq:v12}
\end{equation}
The traditional approach considers two extreme cases, corresponding to 
the somewhat idealised situations of {\sl stable clustering} ($F=1$, a
mean infall streaming that compensates exactly the Hubble flow, such
that clusters are stable in physical coordinates) and {\sl free
expansion} with the Hubble flow ($F=0$, no mean peculiar streaming).  It
is instructive to see explicitly what happens to the contours of $\xi(r_p,\pi)$
in these two limiting cases.  In Figure~\ref{xp-models} I have used
equations \ref{xp_model}, \ref{f_v} and \ref{eq:v12} to plot the model
for $\xi(r_p,\pi)$, 
keeping $\sigma_{12}(r)$ fixed and varying the amplitude $F$ of the
mean streaming.  Here the two competing dynamical 
effects (small--scale stretching and large--scale compression) are
clearly evident.  
The observational results yield values of $\sigma_{12}$ at small
separations around $300-400 \kms$, with a mild dependence on scale
\cite{Fisher_vel,Marzke95, G97}.  This value has been shown to be
rather sensitive to the survey volume, because of the strong
weight the technique puts on galaxy pairs in clusters
\cite{Marzke95}, and the fluctuations in the number of clusters due to
their clustering.   A different method has been proposed more recently
by Landy and collaborators \cite{Landy_xip} to alleviate this
problem.   The method is very elegant, and reduces the weight of
high--velocity pairs in clusters by working in the Fourier domain
where in addition the convolution of the two functions becomes a simple
product of their transforms.  A direct application to data and
n--body simulations under particularly severe survey conditions seems
however to give results which are not significantly dissimilar to the
standard method \cite{Silvia_Marseille}.  

Rather than assuming a model for the mean streaming $v_{12}(r)$, one
could measure it directly from the compression of the contours of
$\xi(r_p,\pi)$, i.e. doing a simultaneous fit to the first and second
moment. This quantity also carries important cosmological information,
being directly proportional to the parameter $\beta=\Omega_M^{0.6}/b $, where 
$\Omega_M$ is the matter density parameter and $b$ is the {\sl bias
parameter} \index{bias}of the class of galaxies one is using (see
Peacock, this volume). 
This has been done, e.g. on the IRAS 1.2Jy survey \cite{Fisher_vel}, but the
uncertainty on $\beta$ is very large due to the weak signal and the
need to simultaneously fit both the first and second moments.  The
situation in this respect will soon improve dramatically thanks to the 
ongoing 2dF \cite{2dF_Dunk} and Sloan (SDSS) surveys \cite{sdss}, that 
will provide 250,000 and 1,000,000 redshifts respectively. 

\section{Is the Universe fractal?}
\label{sec:scaling}
\index{fractal galaxy distribution}

The observation of a power--law shape for the two--point correlation
function together with the self--similar aspect of galaxy maps 
as that of Figure~\ref{2dF-cone}, suggested several years ago a
possible description of the large--scale structure of the Universe in
terms of {\sl fractal objects} \cite{Mandelbrot}.  A fractal Universe
without a cross--over to a homogeneous distribution would
imply abandoning the Cosmological Principle.  Also, under such
conditions most of our standard statistical descriptions of
large--scale structure would be inappropriate 
\cite{Pietronero87}: no mean density could be defined, and as a
consequence the whole concept of {\sl density fluctuations} (with
respect to a mean density) would make little sense.  

It is therefore of significant interest: 1) to compare the scaling
properties of galaxy clustering to those expected for a fractal
distribution (keeping in mind that on different scales there are
different effects at work, as we have seen in the previous section);
2) to put under serious scrutiny the observational evidences for a
convergence of statistical measures to a homogeneus distribution
within the boundaries of current samples.  Attempts to address these
questions using redshift survey data during the last ten years or so
have come to different conclusions, mostly because of disagreement on
which data can be used and how they should be treated and analysed
\cite{Davis_crit,Pietronero_crit,Guzzo97}. It is because of the
relevance of the issues raised that this subject has been the focus of
an intense debate, as also demonstrated by the discussions at this
School (see also Montuori, this book).

\subsection{Scaling laws}

Let us review the arguments for and against the fractal
interpretation of the clustering data, by first recalling the basic
relations involved. 

A fractal set is characterized by a specific 
{\sl scaling} relation, essentially describing the way the set fills
the ambient space.  This scaling law can be by itself taken
as an heuristic definition of fractal (although it is not strictly
equivalent to the formal definition in terms of Hausdorff dimensions,
see e.g. \cite{Provenzale91}): the number of objects counted in spheres of
radius $r$ around a randomly chosen object in the set must scale as
\begin{equation}
N(r) \propto  r^{D}\,\,\,\,\, ,
\label{n_z}
\end{equation}
where $D$ is the {\sl fractal dimension} (or more correctly, the fractal
{\sl correlation} dimension).  \index{fractal dimension}.  Analogously, 
the density within the same sphere will scale as
\begin{equation}
n(r) \propto  r^{D-3}\,\,\,\,\, .
\end{equation}
Similarly, the expectation value of the density measured within
shells of width $dr$ at separation $r$ from an object in the set,
the {\sl conditional density} $\Gamma(r)$ \index{conditional density}
\cite{Pietronero87}, will scale in the same way, 
\begin{equation}
\Gamma(r) = A \cdot  r^{D-3}\,\,\,\,\, ,
\label{gamma}
\end{equation}
with $A$ being constant for a given fractal set.  
$\Gamma(r)$ can be directly connected to the standard two--point
correlation function $\xi(r)$: suppose for a moment that we can define
a mean density $\n_med$ for this sample (we shall see in a moment what
this implies), then it is easy to show that
\begin{equation}
1+\xi(r) = {\Gamma(r) \over \n_med} \propto  r^{D-3}\,\,\,\,\, .
\label{xipiu}
\end{equation}
Therefore, if galaxies are distributed as a fractal, a plot of 
$1+\xi(r)$ will have a power--law shape, and in the strong clustering regime
(where $\xi(r)\gg 1$) this will also be true for the correlation function
itself.  This demonstrates the classic argument (see
e.g. \cite{Peebles80}), that a  
power--law galaxy correlation function as observed $\xi(r) =
(r/r_\circ)^{-\gamma}$, is consistent with a scale--free, fractal
clustering with dimension $D=3-\gamma$ (although it does not
necessarily imply it: fractals are not the only way to produce power--law
correlation functions, see \cite{Guzzo97}).
Note, however, that when $\xi(r)\sim 1$ or smaller, only a plot of
$\Gamma(r)$ or $1+\xi(r)$, 
and not $\xi(r)$, could properly detect a fractal scaling, if present.

When 
this happens over a
range of scales which is significant with respect to the sample size, 
the mean density $\n_med$ becomes an ill--defined 
quantity which depends on the sample size itself.
Considering a spherical sample with radius $R_s$ and the case of a
pure fractal for simplicity, the mean density is the integral of
eq.~\ref{gamma} 
\begin{equation}
\n_med = {3 A \over D} \cdot R_s^{D-3}\,\,\,\,\, ,
\label{meandens}
\end{equation}
and is therefore a function of the sample radius $R_s$.  Under the
same conditions, the two--point correlation function becomes
\begin{equation}
\xi(r) = {\Gamma(r) \over \n_med} - 1 = {D \over 3} \cdot \left({ r 
\over R_s}\right)^{D-3} - 1\,\,\,\,\, ,
\label{xi_rs}
\end{equation}
with a correlation length
\begin{equation}
r_\circ = \left({6 \over D}\right)^{1 \over D-3} \cdot R_s\,\,\,\,\, ,
\label{r0}
\end{equation}
which also depends on the sample size.  
Therefore, if the galaxy distribution has a fractal character,
with a well--defined dimension $D$ one should
observe that: i) The number of objects within volumes of increasing
radius $N(R)$ grows as $R^D$; ii) Analogously, the function
$\Gamma(r)$, or equivalently $1+\xi(r)$, is a power law with slope
$D-3$; iii) The correlation length $r_\circ$ is a linear function of the
sample size. If the fractal distribution extends only up to a certain
scale, the transition to homogeneity would show up first as a
flattening of $1+\xi(r)$ and (less rapidly, given that they depend on
an integral over $r$) as a growth $N(r)\propto r^3$ and a convergency
of $r_\circ$ to a stable value. 

\subsection{Observational evidences}

Pietronero \cite{Pietronero87} made originally the very important point
that the use of $\xi(r)$ was not fully justified, given the size (with 
respect to the clustering scales involved) of the samples available at
the time, and the consequent uncertainty on the value of the mean
density.  In reality, this 
warning was already clear in the original prescription
\cite{Peebles80}: one should be confident to have a {\sl fair
sample} of the Universe before drawing far--reaching conclusions from
the correlation function.  As it often happens, due to the scarcity of
data the recommendation was not followed too strictly (see \cite{Guzzo97}
for more discussion on this point).  
\begin{figure}
\epsfxsize=11cm 
  \hspace{0.2cm}
\epsfbox{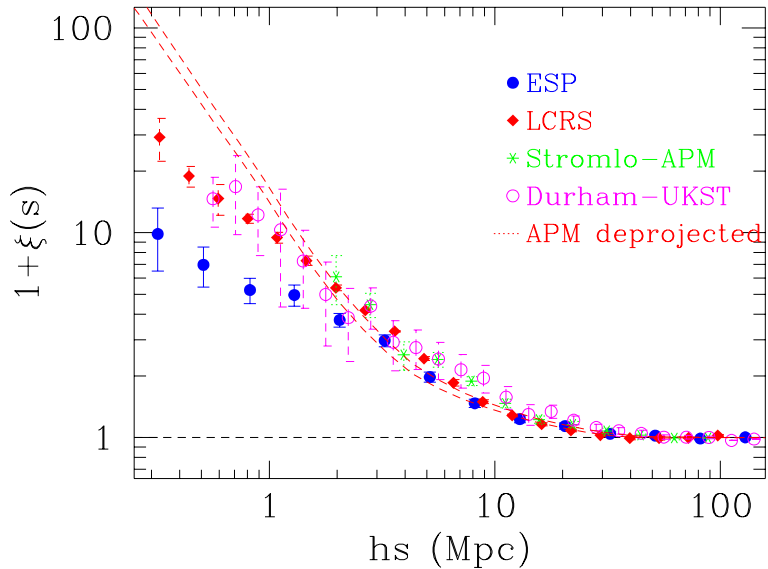} 
\caption{The function $1+\xi(s)$ for the same surveys of
Figure~\ref{xi-surveys}.  A stable power--law scaling would indicate a
fractal range. It is clear how peculiar motions, that affect all data
plotted but the APM $\xi(r)$ which is computed in projection, do distort
significantly the overall shape.  What would seem to be an almost
regular scaling range with $D\sim 2$ from 0.3 to 30 \hmpc, hides in
reality a more complex structure, with a clear inflection around
3\hmpc, which is revealed only when redshift--space effects are eliminated.}
\label{fig-xipiu}
\end{figure}

Although the data available today have increased by an order of magnitude at
least, the debate on the scaling properties and homogeneity of the
Universe is still lively.  Given the subject of these lectures and the
extensive use we have made  
so far of correlation functions, I shall concentrate here on the
evidences concerning points ii) and iii) in the summary list above.
In Figure~\ref{fig-xipiu}, I have plotted the function $1+\xi(s)$ for the
same surveys of Figure~\ref{xi-surveys}.  Taken at face value, the figure
shows that the redshift survey data can be reasonably fitted by a
single power law only out to $\sim 5\hmpc$. However, as soon as we
compare these to the real--space $1+\xi(r)$ from the APM survey, we
realize that what we are seeing here is dominated by the
redshift--space distortions. In other words, a fractal dimension on
small scales  can only be measured from angular or projected
correlations, and if the data are interpreted in this way, it is in
fact close to $D\simeq 1.2$.  Above $\sim 5 \hmpc$, a second range
follows where $D$ varies between 2 and 3, when moving out to scales
approaching $100\hmpc$.  The range between $5 \hmpc$ and $\sim
30\hmpc$ can in principle be described fairly well by a fractal
dimension $D\simeq 2$, as originally found in \cite{G91}, a dimension
that could perhaps be {\sl topological} rather than fractal, reflecting a 
possible sheet--like organisation of structures in this range \cite{paphomo}.
Above $100\hmpc$ the function $1+\xi(r)$ seems to be fairly flat,
indicating a possible convergence to homogeneity.  However, once this
is established, this kind of plot does not allow one to evidence
clustering signals of the order of 1\%, which can only be seen when
the {\sl contrast} with respect to the mean is plotted, i.e. $\xi(s)$.
For a similar analysis and more details, see the pedagogical paper by
Mart\`{\i}nez \cite{Martinez99}.   

Another way of reading the same statistics and on which I would like
to give an update with respect to \cite{Guzzo97} is the scaling of the
correlation length $r_\circ$ with the sample size.   It is 
known that for too small samples there is indeed a growth of $r_\circ$
with the sample size (see e.g. early results in \cite{einasto}).  This
is naturally expected: galaxies are indeed clustered with a power--law
correlation function, and inevitably too small samples will tend
statistically to overestimate the mean density, when measuring it in a
local volume.   When we consider modern samples, however, and we pay
attention to not to compare apples with pears (galaxies with different
morphology and/or different luminosity have different correlation
properties, \cite{Guzzo97}), then the situation is more reassuring:  
Table~\ref{tab-r0} represents an update of that presented in
\cite{Guzzo97}, and reports the general properties of the four redshift
surveys I have used so far as examples.  Being the survey volumes not
spherical, here the ``sample radius'' is defined as that of the maximum sphere 
contained within the survey boundaries (see \cite{Guzzo97}). All these
are estimates of $r_\circ$ in real space.  
\begin{table*}
\begin{center}
\begin{tabular}{lcrcc}  \hline\hline 
Survey & $d$  & $R_s$  & $\ro$ $^{(predicted)}$&
$\ro$ $^{(observed)}$ \\ 
\hline
ESP 		& $\sim 600$	 &  5 &  1.7 & $4.50^{+0.22}_{-0.25}$\\
Durham/UKST 	& $\sim 200$	 & 30 &  10  & $4.6\pm 0.2$\\
LCRS 		& $\sim 400$	 & 32 &  11  & $5.0\pm 0.1$\\
Stromlo/APM 	& $\sim 200$	 & 83 &  28  & $5.1\pm 0.2$\\
\hline
\end{tabular}
\end{center}
\caption{The behaviour of the correlation length $r_\circ$ for the
surveys discussed in previous figures, compared to predictions of a
$D=2$ model.  All estimates 
of $\ro$ are in {\sl real} space.  $d$ is the  
effective depth of the surveys, while the ``sample radius'' $R_s$ has
been computed as in \cite{Guzzo97}.  All measures of distance are expressed in 
$\hmpc$.
}
\label{tab-r0}
\end{table*}                                                                   
The observed correlation lengths are significantly different from the values
predicted by the simple $D=2$ fractal model.  The result would be even
worse using $D=1.2$.  The bare evidence from Table~\ref{tab-r0} is
that the measured values of $\ro$ are remarkably stable, despite
significant changes in the survey volumes and shapes.

The counter--arguments in favour of a fractal interpretation of the
available data are instead summarised in the chapter by M. Montuori.
As the students can check, the main points of disagreement are related
to a) the use of some samples whose incompleteness is very difficult
to assess (as e.g. heterogeneous compilations of data from the 
literature), and b) the estimators used for computing the correlation
function and the way they take the survey shapes into account.  Also
on these issues, the 2dF and SDSS surveys will provide data sets to
fully clear the scene.  In fact, preliminary estimates of the
correlation function from the 2dF survey provide a result in good
agreement with the analyses shown here \cite{2dF_Dunk}.

\subsection{Scaling in Fourier space}
\index{power spectrum}


It is of interest to spend a few words on the complementary, very
important view of clustering in Fourier space. 
The Fourier transform of the correlation function is the power spectrum
$P(k)$
\begin{equation}
P(k) = 4\pi \int_0^\infty \xi(r) {\sin(kr) \over kr} r^2 dr \,\,\,\,\, ,
\end{equation}
which describes the distribution of power among different wavevectors
or {\sl modes} $k=2\pi/\lambda$
once we decompose the fluctuation field $\delta = \delta\rho/\rho$ over
the Fourier basis \cite{JAP}. 
The amount of information contained in $P(k)$ is thus formally the
same yielded by the correlation function, although their estimates 
are differently affected by the uncertainties in the data
(e.g. \cite{JAP,pk_esp}).
\begin{figure}
\epsfxsize=12cm 
  \hspace{0.2cm}
\epsfbox{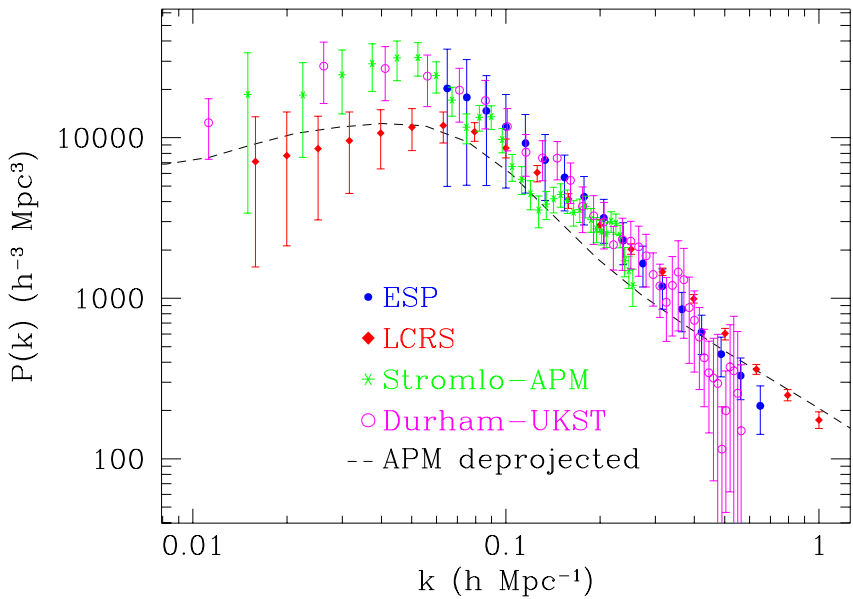} 
\caption{The power spectrum of galaxy clustering estimated from the
same surveys of Figure~\ref{xi-surveys} (also from \cite{Texas}, power 
spectrum estimates from \cite{pk_esp, pk_LCRS, APM-Stromlo-PK, 
Hoyle99}).  Also in Fourier space the differences between real-- and
redshift--space clustering are evident above $k\simeq 0.2\kmpc$.
}
\label{pk_gal}
\end{figure}
One practical benefit of the description of clustering in Fourier
space through $P(k)$\ is that for fluctuations of very
long spatial wavelength ($\lambda > 100 \hmpc$), where \x is
dangerously close to zero and errors easily make the measured values fluctuate
around it (see Figure~\ref{xi-surveys}), $P(k)$ is on the contrary very large.
Around these scales, most models predict a maximum for the power
spectrum, the fingerprint of the size of the horizon at the epoch of
matter--radiation equivalence.  More technical details on power
spectra can be found in the chapter by J. Peacock in this same
book. 

In Figure~\ref{pk_gal}, I have plotted the estimates of $P(k)$\ for the
same surveys of Figure~\ref{xi-surveys}.  Here again the
projected estimate from the APM survey allows us to disentangle the
distortions due to peculiar velocities, which have to be taken
properly into account in the comparisons to cosmological models.  Here
scales are reversed with respect to $\xi(r)$, and the effect manifests
itself in the different slopes 
above $\sim 0.3\kmpc$: an increased
slope in real space (dashed line) corresponds to a stronger damping by
peculiar velocities, diluting the apparent clustering observed in redshift
space (all points).
Below these strongly nonlinear scales, there is a good agreement
between the slopes of the different samples (with the exception of the 
LCRS, see \cite{pk_esp} for discussion), with a well--defined
$k^{-2}$ power law range between $\sim 0.08$ and $\sim  
0.3 \kmpc$.  The APM data show a slope  $\sim
k^{-1.2}$, corresponding to the $\gamma\simeq -1.8$ range of $\xi(r)$,
while at smaller $k$'s (larger scales) they steepen to $\sim k^{-2}$,
in agreement with the redshift--space points. It is this change in
slope that produces the shoulder observed in $\xi(s)$
(cf. \S~\ref{sec:xi}).   Peacock \cite{jap97} showed that such
spectrum is consistent with a steep linear $P(k)$ ($\sim k^{-2.2}$), the 
same value originally suggested to explain the shoulder when first
observed in earlier redshift surveys \cite{G91}.  A dynamical
interpretation of this transition scale has been recently confirmed by a
re--analysis of the APM data \cite{Gatza}.

At even smaller $k$'s all spectra seem to show an indication for a
turnover.  However, when errors are checked in detail, they are at
most consistent with a flattening, with the Durham--UKST survey
providing possibly the cleanest evidence for a maximum around $k\sim
0.03 \kmpc$ or smaller.  A flattening or a turnover to a positive slope 
would be an indication for a scale over which finally the variance is
close to or smaller than that of a random (Poisson) process. But we
learn by looking at older data that a turnover can also be an artifact 
produced when wavelengths comparable to the size of the samples are
considered, and here we are close to that case.

\section{Do we really see homogeneity? Variance on $\sim$1000 h$^{-1}$ 
Mpc scales}
\index{homogeneity}
Wu and collaborators \cite{Wu} and Lahav \cite{Ofer} nicely reviewed
the evidence for a convergence to homogeneity on large scales using
several observational tests.  On scales corresponding to spatial
wavelengths $\lambda \sim 1000 \hmpc$, the 
constraints on the mean--square density fluctuations are provided
essentially by the smoothness in the X--ray and microwave
backgrounds.  Measuring directly the clustering of luminous objects
over such enormous volumes, is only now becoming feasible.  The 2dF
survey will get close to these scales.  The SDSS
\cite{sdss} will do even better through a sub--sample of early
type galaxies selected as to reach a redshift $z\sim 0.5$.  If the
goal of a redshift survey is mapping density fluctuations on the
larges possible scales a viable alternative to using single galaxies
is represented by {\sl clusters of galaxies}.  Here I would like to
discuss the properties of the largest of such surveys, that is in fact 
currently producing remarkable results on the amount of inhomogeneity
on scales nearing 1000 \hmpc.

\subsection{The REFLEX Cluster Survey}
\index{clustering of clusters} \index{REFLEX survey} \index{X-ray clusters}
With mean separations $>10\hmpc$, clusters of galaxies are ideal
objects for sampling efficiently long--wavelength fluctuations over
large volumes of the Universe. 
Furthermore, fluctuations in the cluster distribution are amplified
with respect to those in galaxies, i.e. they are {\sl biased} tracers
of large--scale structure: rich clusters form at the peaks of the
large--scale density field, and their variance is amplified by a factor
that depends on their mass, as it was first shown by Kaiser \cite{Kaiser84}.
X--ray selected clusters have a further major 
advantage over galaxies or other luminous objects when used to trace
and quantify clustering in the Universe: their X--ray emission,
produced through thermal bremsstrahlung by the thin hot
plasma permeating their potential well, is a good measure of their
mass and this allows us to directly compare observations to the
predictions of cosmological models (see \cite{BG2001} for a
review, and \cite{reflex_pk} for a direct application).

\begin{figure}
\epsfxsize=11cm 
  \hspace{0.2cm}
\epsfbox{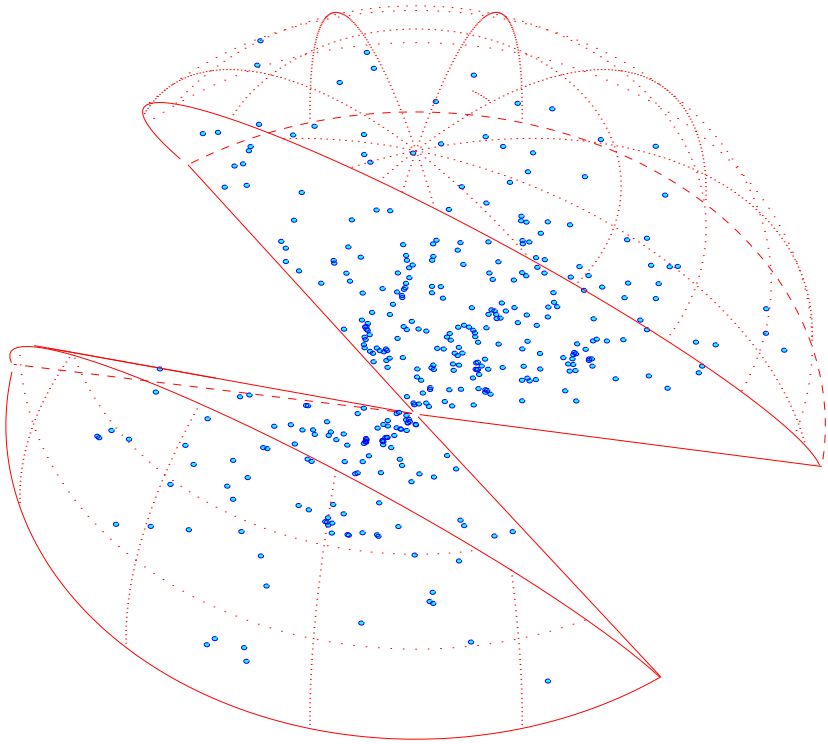} 
\caption{The spatial distribution of X--ray clusters in the REFLEX
survey, out to 600 \hmpc.   Note how, despite the coarser mapping
of large--scale structure, filamentary superclusters (``chains'' of
clusters) are clearly visible. }
 \label{fig:reflex_cone}
 \end{figure}
\begin{figure}
\centering
\epsfxsize=10cm 
  \hspace{0.2cm}
\epsfbox{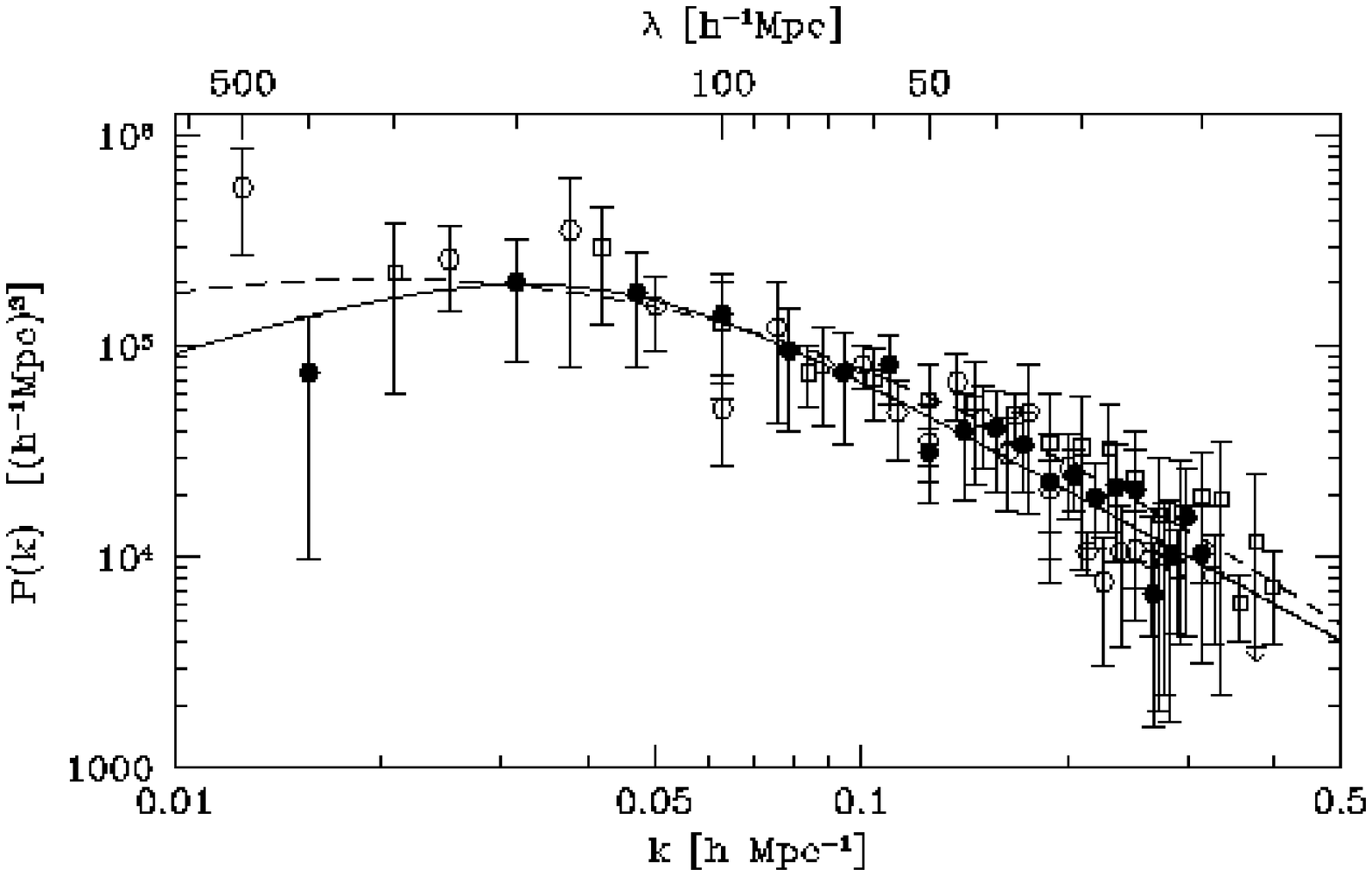} 
%
%
\caption{Estimates of the power spectrum of X--ray clusters from
flux--limited sub--samples of the the REFLEX survey, framed within
Fourier boxes of 300 (open squares), 400 
(filled hexagons), and 500 (open hexagons) h$^{-1}$ Mpc side,
containing 133, 188 and 248 clusters, respectively.  The two lines
correspond to the best--fitting parameters using a phenomenological
shape with two power laws (solid), or a $\Lambda-$CDM model, with
$\Omega_M=0.3$ and $\Omega_\Lambda=0.7$ (dashed) (from
\cite{reflex_pk}). 
}
 \label{fig:pk}
 \end{figure}
The REFLEX (ROSAT-ESO Flux Limited X-ray) cluster survey is the result
of the most intensive effort for a homogeneous identification of
clusters of galaxies in the ROSAT All Sky Survey (RASS). It combines
a thorough analysis of the X--ray data , and extensive optical
follow--up with ESO telescopes, to construct a complete  
flux--limited sample of about 700 clusters with measured redshifts and 
X-ray luminosities \cite{Hans98, gigi99}.   The survey covers most of the southern celestial 
hemisphere ($\delta<2.5^\circ$), at galactic latitude 
$|b_{II}|>20^\circ$ to avoid high absorption and stellar crowding.  
The present, fully identified version of the REFLEX survey contains
452 clusters 
and is more than 90\% complete to a nominal flux limit of 
$3 \times
10^{-12}$ erg s$^{-1}$ cm$^{-2}$ (in the ROSAT band, 0.1--2.4
keV). Mean redshifts for virtually all these have been measured during a long
observing campaign with ESO telescopes.  Details on the 
identification procedure and the survey properties can be found in
\cite{survey_paper}, while earlier results are reported in
\cite{DeGrandi_XLF,RASS1}. 

Figure~\ref{fig:reflex_cone} shows the spatial distribution 
of REFLEX clusters, evidencing a number of superstructures with sizes
$\sim 100 \hmpc$.  One of the main motivations for this survey was
to compute the power spectrum on extremely large scales, benefiting of
the efficiency of cluster samples to cover very large volumes of the
Universe.  Figure~\ref{fig:pk} shows the estimates of $P(k)$ from three
subsamples of the survey (from \cite{reflex_pk}).  

\begin{figure}
\epsfxsize=12cm 
  \hspace{0.2cm}
\epsfbox{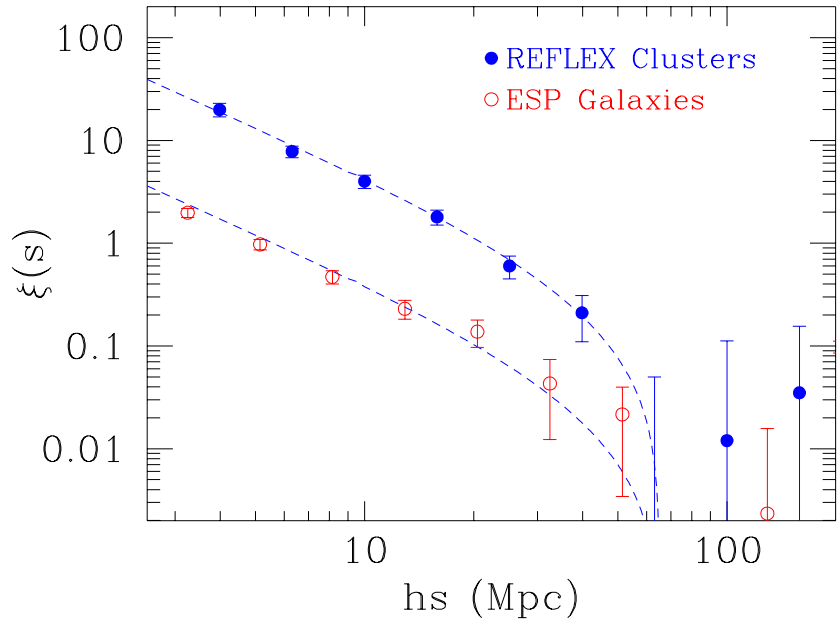} 
\caption{The two--point correlation function of the whole
flux--limited REFLEX cluster catalogue (filled circles,
\cite{xi_reflex}), compared to 
that of ESP galaxies (open circles, \cite{GuzzoESP}. 
The dashed lines show the Fourier transform of a phenomenological fit
to $P(k)$ which tries to include the large--scale power seen from the
largest subsamples (top line).  The bottom curve is that obtained
after scaling down by an arbitrary bias factor ($b_c^2=(3.3)^2$ in
this specific case). 
}
\label{fig:xi}
\end{figure}
One of the strong advantages of working with X--ray selected clusters of
galaxies is that connection to model predictions is far less ambiguous 
than with optically--selected clusters (e.g. \cite{Moscardini2000,
BG2001}).  We have therefore used the specific REFLEX 
selection function (converted essentially to a selection in mass), to
determine that a low--$\Omega_M$ model (open or $\Lambda$--dominated),
best matches {\bf both} the shape and amplitude (i.e. bias value) of
the observed power spectrum \cite{reflex_pk} (dashed line in the figure).  
In fact, the samples shown here do not reach the maximum spatial
wavelengths we can possibly sample with 
the current data, as the Fourier box could be made to be as large as
1000 h$^{-1}$ Mpc (the survey reaches $z=0.3$ with the most luminous
objects).  In such case, however, our control over systematic
effects becomes poorer, and work is currently undergoing to pin errors 
down and understand how trustable are our results on $\sim 1$ Gpc
scale, where we do see extra power coming up.  At the very least,
REFLEX is definitely showing more clustering power on very large
scales than any galaxy redshift survey to date.  Similar hints for
large--scale inhomogeneities seem to be suggested by the most recent
analysis of Abell--ACO samples \cite{Miller}.  

For $k>0.05 \kmpc$, on the other hand, a comparison of REFLEX to
galaxy power spectra shows a rather similar shape. This is probably
better appreciated by looking at the two--point correlation  
function $\xi(s)$ \cite{xi_reflex}, compared in Figure~\ref{fig:xi} to that
of the ESP galaxy redshift survey.   The agreement in shape between
galaxies and clusters is remarkable on all scales, with a break to zero around
$60-70\hmpc$ for both classes of objects.  This is in general expected
in a simple biasing scenario where clusters represent \index{bias}
the high, rare peaks of the mass density distribution. This result
strongly corroborates the simpler, reassuring view that 
at least above $\sim 5 \hmpc$ the galaxy and mass distributions are
linked by a simple constant bias.

\subsection{``Peaks and Valleys'' in the Power Spectrum}
\index{power spectrum} \index{high--baryon models}
Most of the discussion so far has been concentrating on the beauty of finding
``smooth'' simple shapes for $\xi(r)$ or $P(k)$, as symptoms of an
underlying order of Nature.  Rather than being a demonstration of Nature
inclination for elegance, however, this smoothness and simplicity might
simply indicate our ignorance and lack of data.  In fact, while smooth
power spectra 
are predicted in models dominated by non--interacting dark matter
particles, as Cold Dark Matter, a very different situation is expected
in cases where ordinary (baryonic) matter plays a more significant
role, with wiggles appearing in $P(k)$ that would be difficult to detect
with the size and ``Fourier resolution'' of our current data sets.

The possibility that the power spectrum shows a sharp
peak (or more peaks) around its maximum has been suggested a few times
during the last few years. 
For example, Einasto and collaborators \cite{Einasto_peak} found
evidence for a sharp peak around $k\simeq 0.05\kmpc$ in the power
spectrum of an earlier sample of Abell clusters, a feature later
confirmed with lower significance by a more conservative analysis of
the same data \cite{Retzlaff98}.  The position of this feature is
remarkably close to the $\sim 130\hmpc$ ``periodicity'' revealed
by Broadhurst and collaborators in a ``pencil--beam'' survey towards
the galactic poles \cite{BEKS}, and more recently, in an analysis of
the redshift distribution of Lyman--break selected galaxies
\cite{Lyb_peaks}. Other evidences have been claimed from
two--dimensional analyses of redshift ``slices'' \cite{Landy_2D}, or
QSO superstructures \cite{Boud}.

These observations have stimulated some interesting work on
models with high baryonic content.  In this case, the power spectrum
can exhibit a detectable inprint from ``acoustic'' oscillations within
the last scattering surface at $z\sim 1000$, the same features
observed in the Cosmic Microwave Background (CMB) radiation
(\cite{Eisenstein}).  While the most recent estimates of the REFLEX
power spectrum do not show clear features around the scales of
interest to justify ``extreme'' high--baryon models (contrary to early
indications \cite{Guzzo_Dunk}, which shows the importance of the careful
assessment of errors), the extra power below $k\sim 0.02$ could still
be an indication of an higher--than--conventional baryon fraction
\cite{Eisenstein, reflex_lss}, along the lines that seem to be
suggested by the Boomerang CMB results \cite{Boomerang}.

\section{Conclusions}

At the end of these lectures, a student is possibly more confused
than he/she was in the beginning, at least after a first read.  I hope 
however that once the dust settles, a few important points emerge.
First, that the processes which shaped the large--scale distribution
of luminous objects we observe today are different at different
scales. At small scales, we observe essentially the outcome of fully
nonlinear gravitational evolution that re--shaped the linear power
spectrum into a collection of virialised, or nearly so structures.
Therefore, one cannot naively take the redshift survey data and look
for specific patterns or statistical properties without taking 
into account galaxy peculiar motions.  For this reason, one should be
careful in over--interpreting things like a single power--law scaling
from scales of a tenth of a Megaparsecs to hundred Megaparsecs,
because, again, different phenomena are being compared. On the
contrary, one can use these distortions to really ``see'' how the true
mass distribution is, and I have spent a considerable part of these
lectures to describe some of the techniques in use.  
 Moving to larger and larger scales, we enter a regime where we are
lucky enough that we can still see something related to the original
scaling law of fluctuations. This is what was originally produced by
some generator in the early Universe (inflation?) and processed
through a matter (dark plus baryons) controlled amplifier. On even
larger scales, we hope we are finally entering a regime where the
variance in the mass is consistent with a homogeneous
distribution, although we have seen that even the largest galaxy and
cluster samples are barely sufficient to see hints of that, perhaps
suggesting even more inhomogeneity than we expect.  Does this mean
that we are living in a pure fractal Universe? The scaling behaviour
of galaxies and the stability of the correlation length seem to imply
that this cannot be the case.  On top of everything, the smoothness of 
the Cosmic Microwave Background (treated elsewhere in this book) is
probably the most reassuring observation in this respect.  
What we seem to understand is that our samples have still difficulty
to properly sample the very largest fluctuations of the density field, 
on scales where this is not fully Poissonian (or sub--Poissonian) yet. 

Finally, I hope the students get the message that despite the
tremendous progress of the last 25 years which transformed Cosmology
into a real science, we still have a number of fascinating questions
to answer and still feel far away from convincing ourselves that we
have understood the Universe.

\section*{Acknowledgments}
Most of the results I have shown here rely upon the work of a number of
collaborators in different projects. I would like to thank in
particular my colleagues in the REFLEX collaboration, especially
C. Collins and P. Schuecker for the work on correlations and power
spectra shown here.  Thanks are due to F. Governato for providing me
with the simulation used for producing Figure~\ref{sim-cones}, and to
Alberto Fernandez--Soto  and Davide Rizzo for a careful reading of the
manuscript. Finally, thanks are due to the organizers of the Como
School, for their patience in waiting for this paper and for allowing
me extra page space.

\Bibliography

\end{document}